\documentclass{aa}
\usepackage{txfonts}
\usepackage{natbib}
\usepackage{graphicx}
\bibpunct{(}{)}{;}{a}{}{,}

\begin{document}

\newcommand{\degree}{\ensuremath{^{\circ}}}
\newcommand{\referenssi}{\fbox{Referenssi!}}
\newcommand{\work}{\fbox{WORK!}}
\newcommand{\sg}{$\sigma$ Gem}
\newcommand*\fy{\ensuremath{\overset{\text{y}}{.}}}

\newcommand{\kuiperperiod}{\ensuremath{P=19.6033398395}}
\newcommand{\orbitalperiod}{\ensuremath{P_{\rm{orb}}=19.6}}
\newcommand{\pind}{\ensuremath{19\fd50 \pm 0\fd37}}
\newcommand{\Pmag}{\ensuremath{P_{\rm{M}}=6.69\pm0.21}}
\newcommand{\Pamp}{\ensuremath{P_{\rm{A}}=3.12\pm0.25}}
\newcommand{\Pper}{\ensuremath{P_{\rm{P}}=4.4\pm1.0}}
\newcommand{\chimag}{\ensuremath{\chi_{\rm{red}}^2 = 399.2}}
\newcommand{\chiamp}{\ensuremath{\chi_{\rm{red}}^2=29.2}}
\newcommand{\chiper}{\ensuremath{\chi_{\rm{red}}^2=5.92}}

\newcommand{\Zval}{\ensuremath{Z=0.103}}
\newcommand{\hrange}{\ensuremath{-}}
\newcommand{\krange}{\ensuremath{0.14 \la \alpha \la 0.21}}
\newcommand{\kmin}{\ensuremath{0.14}}
\newcommand{\kmax}{\ensuremath{0.21}}
\newcommand{\delorange}{\ensuremath{0.05-0.07}}
\newcommand{\Pmean}{\ensuremath{P_{M}=15.1\pm} 0.49}

\newcommand{\kppri}{$P_{\rm{min},1}=19\fd472405\pm 0\fd000020$}
\newcommand{\kppriQ}{$Q=1.05\times 10^{-6}$}
\newcommand{\kpsec}{$P_{\rm{min},2}=19\fd635$, $Q=8.27\times 10^{-6}$}
\newcommand{\kpboth}{$P_{\rm{min},1,2}=19\fd6040216\pm 0\fd0000051$}
\newcommand{\kpbothQ}{$Q=8.52\times 10^{-6}$}

\newcommand{\rhoAP}{\ensuremath{\rho = -0.18}} 
\newcommand{\pvalAP}{\ensuremath{p=0.13}}
\newcommand{\rhoMA}{\ensuremath{\rho = 0.33}} 
\newcommand{\pvalMA}{\ensuremath{p=0.004}}
\newcommand{\rhoMP}{\ensuremath{\rho = -0.07}} 
\newcommand{\pvalMP}{\ensuremath{p=0.58}}

\newcommand{\ff}{ff-event}
\newcommand{\ab}{ab-event}
\newcommand{\gr}{gr-event}

\newcommand{\spLambdaI}{108\degree}
\newcommand{\spLambdaII}{284\degree}
\newcommand{\spRI}{25\degree}
\newcommand{\spRII}{25\degree}
\newcommand{\spBetaI}{20\degree}
\newcommand{\spBetaII}{10\degree}
\newcommand{\spU}{0.79}
\newcommand{\spKappa}{0.57}
\newcommand{\spZ}{0.033}

\title{Spot activity of the RS CVn star $\sigma$ Geminorum \thanks{The analysed photometry and numerical results of the analysis are both published electronically at the CDS via anonymous ftp to cdsarc.u-strasbg.fr (130.79.128.5) or via http://cdsarc.u-strasbg.fr/vizbin/qcat?J/A+A/yyy/Axxx}}

\author{P. Kajatkari\inst{\ref{inst1}}
\and T. Hackman\inst{\ref{inst2}}
\and L. Jetsu\inst{\ref{inst1}}
\and J. Lehtinen\inst{\ref{inst1}}
\and G.W. Henry\inst{\ref{inst3}}}

\institute{Department of Physics, P.O.Box 64, FIN-00014 University of Helsinki, Finland\label{inst1}
\and
Finnish Centre for Astronomy with ESO (FINCA), University of Turku, Väisäläntie 20, FI-21500 Piikkiö, Finland\label{inst2}
\and
Center of Excellence in Information Systems, Tennessee State University,\\ 3500 John A. Merritt Blvd., Box 9501, Nashville, TN 37209, USA\label{inst3}}
\date{Received date / Accepted date}

\abstract{}
{We model the photometry of RS CVn star \object{$\sigma$ Geminorum} to obtain 
new information on the changes of the surface starspot distribution, i.e., activity cycles, differential rotation and active longitudes.}
{We use the previously published Continuous Periods Search-method (CPS) to analyse V-band differential photometry obtained between the years 1987 and 2010 with the T3 0.4 m Automated Telescope at the Fairborn Observatory. The CPS-method divides data into short subsets and then models the light curves with Fourier-models of variable orders and provides estimates of the mean magnitude, amplitude, period and light curve minima. These light curve parameters are then analysed for signs of activity cycles, differential rotation and active longitudes.}
{We confirm the presence of two previously found stable active longitudes, synchronised with the orbital period $P_{\mathrm{orb}}=19\fd60$ and find eight events where the active longitudes are disrupted. The epochs of the primary light curve minima rotate with a shorter period $P_{\mathrm{min,1}}=19\fd47$ than the orbital motion. If the variations in the photometric rotation period were to be caused by differential rotation, this would give a differential rotation coefficient of $\alpha \ge 0.103$. }
{The presence of two slightly different periods of active regions may indicate a superposition of two dynamo modes, one stationary in the orbital frame and the other one propagating in the azimuthal direction. Our estimate of the differential rotation is much higher than previous results. However, simulations show that this can be caused by insufficient sampling in our data.}


\maketitle
\section{Introduction}

The RS CVn-type star \object{$\sigma$ Geminorum}
is a bright ($\rm{V} \approx 4.14$), 
variable binary 
with a relatively long orbital period $P_{\rm{orb}}= 19\fd604471$ \citep{duemmler1997}. The primary component is a K1III giant, but the secondary is not visible and has no noticeable effect on the spectrum of the binary. The secondary is most likely a cool, low-mass main-sequence star or possibly a neutron star \citep{duemmler1997, Ayres1984}. The inclination of the rotational axis of the primary is roughly $60 \degree$ \citep{eker1986}.

The photometric variability of \sg{} was first detected by \citet{hall1977}. Since 1983, intensive and continuous photometric observations have been made with automated photometric telescopes (APT). The light curves acquired in this fashion have been studied in detail, e.g. by \citet{fried1983}, \citet{henry1995}, \citet{jetsu1996} and \citet{zhang1999}.

Doppler imaging has been used to construct surface temperature maps of \sg{} \citep{hatzes1993,kovari2001}. These surface images had no polar spots, a feature often reported in other active stars. Instead, the spot activity appears to be constrained into a latitude band  between $30\degree$ and $60\degree$. 

In most late-type stars, no unique, regular and persistent activity cycle has been found. In the case of \sg{}, various analyses have yielded a wide range of different possible 
quasi-periodicities, which are assumed to be an indication of a possible stellar cycle, similar to the 11-year sunspot cycle. \citet{strassmeier1988} suggested a possible 2.7 year period in the spotted area of \sg. \citet{henry1995}, who found a cycle of 8.5 years instead, suggested that the 2.7 year period is related to the lifetime of individual spot regions and hence this shorter period would not represent a true spot cycle. They also attributed the 5.8 year cycle found by \citet{maceroni1990} to the spot migration rate determined by \citet{fried1983}.

The light curve minima of \sg{} have shown remarkable stability in phase over time span of years, or even decades. This indicates a presence of active longitudes, a phenomenon often seen in chromospherically active stars. Active longitudes are longitudinally concentrated areas that show persistent activity, manifesting as starspots. Active longitudes on \sg{} have previously been studied by \citet{jetsu1996} and \citet{berdyugina1998}. The results indicate that the active longitudes are synchronised with the orbital period, with a preference to the line connecting the binary components. \citet{berdyugina1998} also suggested that there is a possible 14.9 year activity cycle in the star.

Differential rotation has been studied using photometric spot models and Doppler-imaging techniques. \citet{henry1995} used spot modelling to determine the migration rate of the starspots and arrived at a value 
for the differential rotation coefficient. \citet{kovari2007c} analysed Doppler images, using the Local correlation technique (LCT). Their analysis indicated anti-solar differential rotation with $\alpha=-0.0022\pm0.0016$. Using a different technique for the same data, they also got another value $\alpha=-0.0021\pm0.005$ 
\citep{kovari2007b}.

\section{Observations}

The observations in this paper are differential photometry in the Johnson $V$ passband obtained at Fairborn Observatory in Arizona using the 0.4 m T3 Automated Photometric Telescope (APT). Each observation is a sequence of measurements that were taken in the following order: K, sky, C, V, C, V, C, V, C, sky and K, where K is the check star, C is the comparison star and V the program star. The comparison star was HR 2896 and the check star was $\upsilon$ Gem. Until 1992, the precision of the measurements was 0.012 mag. Then a new precision photometer was installed and the precision of the subsequent measurements has been $\sim 0.004-0.005$ mag \citep{fekel2005}. A thorough description of the APT observing procedures has been given by \citet{henry1999}.

\begin{figure*}
\centering
\includegraphics[width=17cm]{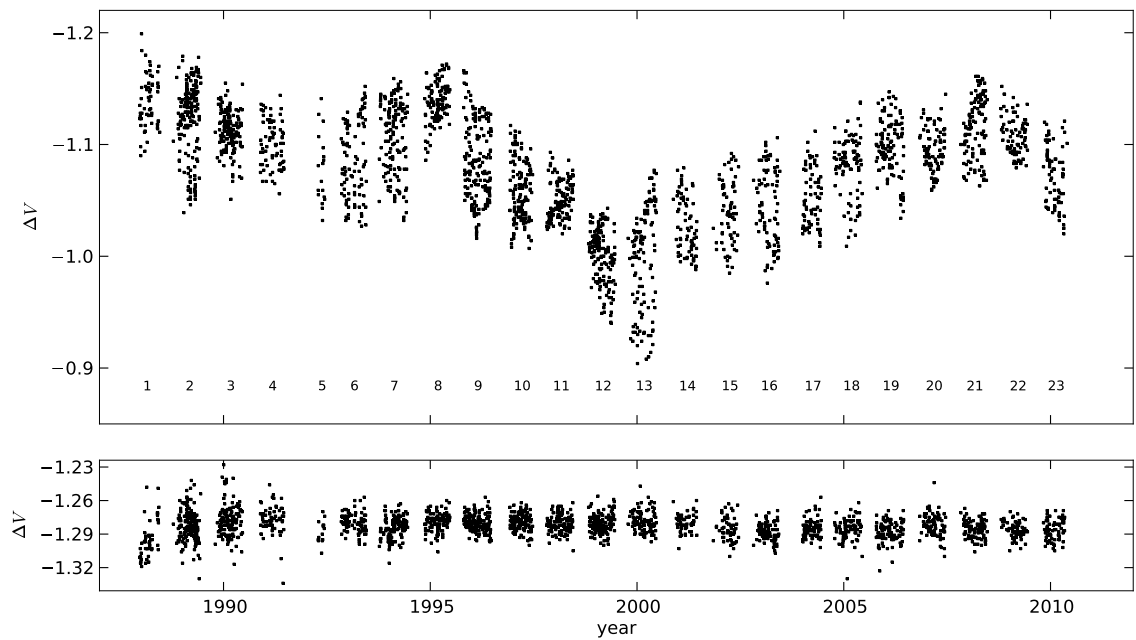}
\caption{The differential magnitudes between \sg{} and HR 2896, and the check star $\upsilon$ Gem and HR 2896 between the years 1987 and 2010. Different observing seasons are denoted by their corresponding segment number. }
\label{raw_data}
\end{figure*}

The whole time series consists of 2683 observations and spans from JD 2447121.0481 (21 November 1987) to 2455311.6556 (25 April 2010). The V-C and K-C differential magnitudes are shown in Fig. \ref{raw_data}. The numbers displayed in the upper panel refer to the segment division and correspond to different observing seasons. We decided against including previously published data from other sources. The continuous period search method (hereafter CPS)is best suited for temporally continuous data of homogeneous quality and inclusion of temporally sparse data may induce unreliable results. This is also the approach taken in earlier studies utilising the CPS, i.e., \citet{lehtinen2012, hackman2013}.

\section{Data analysis}

Here we give a short introduction to the CPS-method and how we used it in the time series analysis of our paper. A complete description of the method can be found in \citet{lehtinen2011}. The CPS-method has been developed from the Three Stage Period Analysis (TSPA) by \citet{jetsu1999}. The CPS uses a sliding window to divide the data into shorter datasets and then determines local models using a variable $K$th-order Fourier series:
\begin{equation}
\hat{y}(t_i) = \hat{y}(t_i,\bar{\beta}) = M + \sum_{k=1}^K{[B_k\cos{(k2\pi ft_i)} + C_k\sin{(k2\pi ft_i)}]}.
\label{model}
\end{equation}

\noindent The optimal model order $K$ used for each dataset is determined by the Bayesian information criterion. The highest modelling order used in this study was $K=2$. The possibility of a constant model $K=0$ is also considered. In that case, the model is simply the weighted mean of the data points $y_{i} = y(t_{i})$ in the dataset.

The first step of the CPS-analysis is to divide the data into datasets. The datasets are composed using a rectangular window function with a predetermined length $\Delta T_{\rm{max}}$ that is moved forward through the data one night at a time. A new dataset is created each time when the dataset candidate determined by the window function includes at least one new data point that was not included in the previous dataset. Each modelled dataset must also include at least $n_{\rm{min}}$ data points to be valid. We used values $n_{\rm{min}}=14$ and which is roughly two and half times the average photometric period. 

The first dataset with a reliable model is called an {\it independent} dataset. The next independent datasets are selected with the following two criteria. Firstly, this next independent dataset must not share any common data with the previous independent dataset. Secondly, the model for this next independent dataset needs to be reliable. In other words, these independent datasets do not overlap and their models are always reliable. With this definition, the correlations between the model parameters of independent datasets represent real physical correlations, i.e. these correlation are not due to bias caused by common data.

The datasets are combined into segments, each representing a different observing season. The segment division does not directly affect the analysis because each dataset is still analysed separately. The segment division of this analysis is given in Table \ref{segments}. For each segment, the length of the segment is given, along with the total number of data points, the number of datasets and number of independent datasets. 

The parameters obtained from the light curve model, as a function of the mean epoch of the dataset, $\tau$, are:
\begin{itemize}
\item[]{$M(\tau)=$ mean magnitude}
\item[]{$A(\tau)=$ peak to peak light curve amplitude}
\item[]{$P(\tau)=$ photometric period}
\item[]{$t_{\rm{min,1}}(\tau)=$ epoch of the primary minimum}
\item[]{$t_{\rm{min,2}}(\tau)=$ epoch of the secondary minimum}
\item[]{$T_{\rm{C}}(\tau)=$ time scale of change.}
\end{itemize}

The CPS also provides graphical representation of the results for each segment. An example of this is given later in Fig.\ref{seg3}. That figure contains the following panels:

\begin{itemize}
\item[](a) standard deviation of residuals $\sigma(\tau)$;
\item[](b) modelling order $K(\tau)$ (squares, units on the left $y$-axis); and
number of observations per dataset $n$ (crosses, units on the right $y$-axis);
\item[](c) mean differential $V$-magnitude $M(\tau)$;
\item[](d) time scale of change $T_{\rm{C}}(\tau)$
\item[](e) amplitude $A(\tau)$;
\item[](f) period $P(\tau)$;
\item[](g) primary (squares) and secondary (triangles) minimum phases $\phi_{\rm{min},1}(\tau)$ and $\phi_{\rm{min},2}(\tau)$;
\item[](h) $M(\tau)$ versus $P(\tau)$;
\item[](i) $A(\tau)$ versus $P(\tau)$;
\item[](j) $M(\tau)$ versus $A(\tau)$;
\end{itemize}

Reliable models are denoted with closed symbols and unreliable with open symbols.

\begin{figure*}
\centering
\includegraphics[width=17cm,angle=-90]{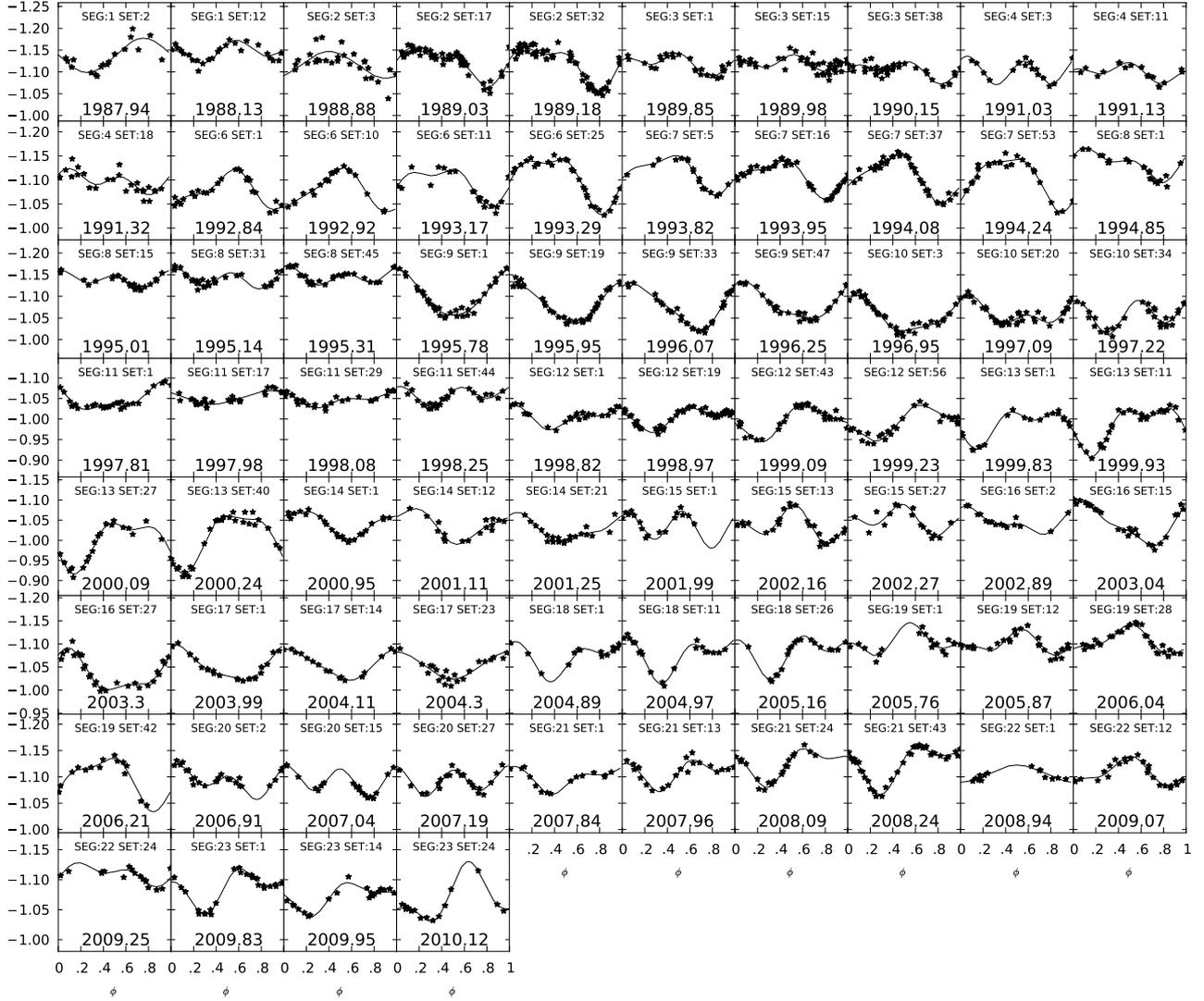}
\caption{Light curves and the best-fit models of independent datasets. The procedure used to calculate the phases is explained at the end of Section 3.}
\label{lightcurves}
\end{figure*}

The numerical results of the CPS analysis can be accessed electronically at the CDS. The light curves and the best-fit models of the independent datasets are shown in Fig. \ref{lightcurves}. The light curves are plotted as a function of phase $\phi=\phi_{i}+\phi^{\prime}_{i}$. For each dataset, the phases $\phi_{i}$ were first calculated using the best-fit periods $P(\tau)$ and the epochs of the primary minima $t_{\rm{min},1}(\tau)$. The phases of each dataset were then adjusted by $\phi^{\prime}=\phi_{\rm{orb},1}-0.2$, where $\phi_{\rm{orb},1}$ are the phases of the primary minimum epochs $t_{\rm{min},1}$ of each dataset, calculated using the orbital ephemeris $\rm{JD}_{\rm{conj}}= 2447237\fd02 +19\fd604471 E$. This adjustment was done in order to get the phases of the primary minima to be the same as in Fig. \ref{phase_diag}. Similar procedure was done in \citet{lehtinen2011}.

Some of the light curves in segments 2, 9 and 17 show clearly that the brightness of the star can change during two successive rotations. This could have partly been avoided by using a shorter window. We also analysed the data using a window $\Delta T_{\rm{max}} =39\fd2$, but the result was a large number of unreliable models, thus the longer window $\Delta T_{\rm{max}} =49\fd0$ was used in the final analysis. 

\begin{table}
\caption{Segments of the \sg{} photometry. Columns from left to right are: Segment number, observing time interval, number of data points, total number of datasets and number of independent datasets. }

\centering
\begin{tabular}{clccc}
\hline
\hline
SEG & Interval & n & sets & ind. sets\\
\hline
1 & 21. 11. 1987 - 11. 3. 1988 & 47 & 13 & 2 \\
2 & 13. 10. 1988 - 13. 5. 1989 & 182 & 49 & 3 \\
3 & 6. 10. 1989 - 15. 5. 1990 & 155 & 51 & 3 \\
4 & 22. 10. 1990 - 16. 5. 1991 & 77 & 18 & 3 \\
5 & 14. 3. 1992 - 6. 5. 1992 & 17 & 3 & 0 \\
6 & 5. 10. 1992 - 13. 5. 1993 & 97 & 27 & 4 \\
7 & 5. 9. 1993 - 13. 5. 1994 & 142 & 59 & 4 \\
8 & 12. 10. 1994 - 21. 5. 1995 & 119 & 47 & 4 \\
9 & 21. 9. 1995 - 19. 5. 1996 & 170 & 59 & 4 \\
10 & 3. 11. 1996 - 21. 5. 1997 & 138 & 46 & 3 \\
11 & 26. 9. 1997 - 15. 5. 1998 & 132 & 55 & 4 \\
12 & 29. 9. 1998 - 20. 5. 1999 & 152 & 65 & 4 \\
13 & 28. 9. 1999 - 17. 5. 2000 & 119 & 49 & 4 \\
14 & 12. 11. 2000 - 11. 5. 2001 & 72 & 23 & 3 \\
15 & 26. 11. 2001 - 14. 5. 2002 & 63 & 27 & 3 \\
16 & 24. 10. 2002 - 14. 5. 2003 & 82 & 27 & 3 \\
17 & 3. 12. 2003 - 13. 5. 2004 & 68 & 24 & 3 \\
18 & 19. 10. 2004 - 9. 5. 2005 & 84 & 39 & 3 \\
19 & 13. 9. 2005 - 11. 5. 2006 & 114 & 48 & 4 \\
20 & 16. 10. 2006 - 17. 5. 2007 & 82 & 39 & 3 \\
21 & 30. 9. 2007 - 18. 5. 2008 & 114 & 56 & 4 \\
22 & 11. 11. 2008 - 9. 5. 2009 & 69 & 26 & 3 \\
23 & 29. 9. 2009 - 24. 4. 2010 & 78 & 28 & 3 \\
\hline
\end{tabular}

\label{segments}
\end{table}

\section{Results}

\subsection{Long term variability and activity cycles}

\begin{figure*}
\centering
\includegraphics[width=17cm]{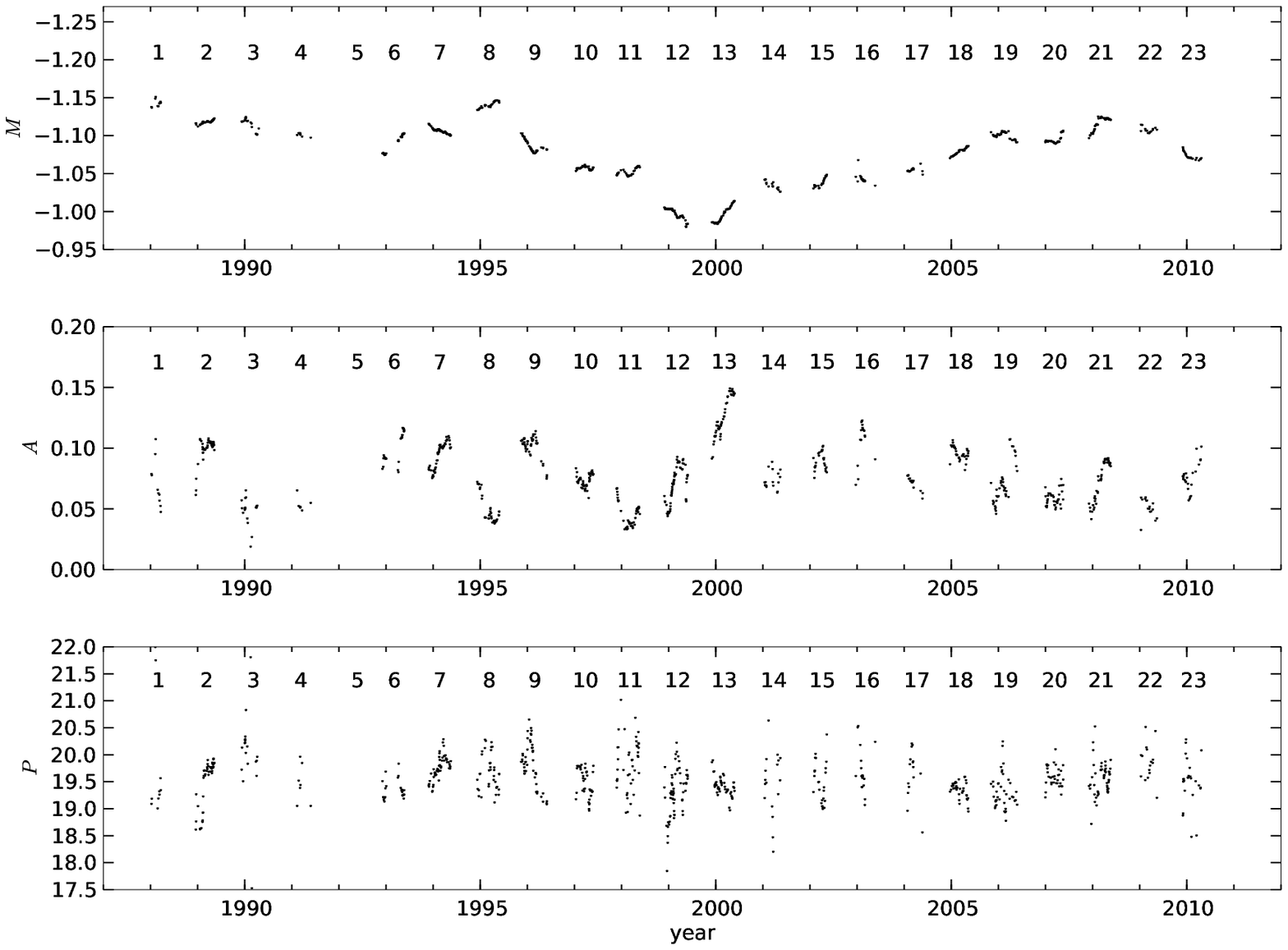}
\caption{The long-term changes of mean ($M$), amplitude ($A$) and period ($P$). The segment numbers are shown above the data.}
\label{amp}
\end{figure*}

The photometry of \sg{} has been studied before with intention of searching for long-term activity cycles, but so far, none of the findings has been conclusive. We applied the CPS to the $M$, $A$ and $P$ estimates of independent datasets, using a first-order model ($K=1$). The long-term changes of these parameters are shown in Fig. \ref{amp}. For the mean magnitudes $M$, we find the best period \Pmag{} yr. For $A$ the best period was \Pamp{} yr and for $P$, the best period was \Pper{} yr. 

We also checked the parameters $A$, $M$ and $P$ from independent datasets for correlations. One might expect a correlation between the mean brightness and starspot amplitude, simply because when larger parts of the star are covered by starspots, the star should appear dimmer. The starspot amplitude and period might also correlate, with changing latitude the effective covered area seen by the observer changes and due to differential rotation, if present, the period might also change.

As the dependencies between these parameters are not necessarily linear, we calculated the Spearman's rank correlation coefficient and the corresponding p-value for each pair of parameters. Unsurprisingly, we find the strongest correlation between $M$ and $A$, with a correlation coefficient \rhoMA{} and a p-value \pvalMA{}. For $M$ and $P$ we get \rhoMP{} and \pvalMP{}, and for $P$ and $A$, \rhoAP{} and \pvalAP{}, none of which is statistically significant. Although there is relatively strong correlation between $A$ and $M$, the periods  $P_\mathrm{A}$ and $P_\mathrm{M}$ are different.  This is at least partly explained by the spot coverage being sometimes more axisymmetric, like in the Doppler images by \citet{kovari2001}. In the light curves obtained during similar spot configurations, the peak-to-peak amplitude of the light curve can be low, even though the star would appear to be faint.

\subsection{Differential rotation}

In order to estimate the stellar differential rotation, starspots have been used as markers that are assumed to rotate across the visible stellar disc with varying angular velocities, determined by their respective latitudes. To estimate the amount of surface differential rotation present in the star, we use the dimensionless parameter 
\begin{equation}
Z=\frac{6\Delta P_{\rm{W}}}{P_{\rm{W}}},
\label{z_parameter}
\end{equation}
where $P_{\rm{w}}\pm\Delta P_{\rm{w}}$ is the weighted average of periods from the independent datasets $P_{i}$, given by $P_{\rm W } = (\Sigma w_i P_i)/\Sigma P_i$, $\Delta P_{\rm w} = \sqrt{\Sigma w_i(P_i - P_{\rm{w}})^2/\Sigma w_i}$ and $w_i = \sigma_{\rm{P}}^{-2}$ \citep{jetsu1993a}. 
The parameter $Z$ gives the $\pm3\Delta P_{\rm w}$ upper limit for the variation of the photometric period $P_{\rm{phot}}$. 

Using only the period estimates from the independent datasets, we get the weighted mean of the photometric period $P_{\rm{w}}\pm\Delta P_{\rm{w}}=\pind$, which gives \Zval{}. Our amplitude to noise ratio for the typical light curve amplitude $A(\tau)=0.10$ mag was about 100. This means that spurious changes of $Z$ caused by noise were not significant \citep[][Table 2]{lehtinen2011}. 

We can use the parameter $Z$ to derive the differential rotation profile of a star (assuming Solar-like differential rotation) 
\begin{equation}
\Omega(l)=\Omega_{0}(1-\alpha \sin^2(l))
\label{rot_profile}
\end{equation}
where $l$ is the latitude, $\Omega_{0}$ is the rotation rate at the equator and $\alpha$ the differential rotation coefficient. The value of $\alpha$ can be 
 estimated with the relation $\vert \alpha \vert \approx Z/h$
\citep{jetsu2000a}, where $h=\sin^{2}(l_{\rm{max}})-\sin^{2}(l_{\rm{min}})$, and the parameters $l_{\rm{min}}$ and $l_{\rm{max}}$ are the minimum and maximum latitudes between which the spot activity is confined.

Doppler imaging results by \citet{hatzes1993} and \citet{kovari2001} indicate that the most of the spot activity on \sg{} is constrained to latitudes between $30\degree$ and $60\degree$, with some activity on lower latitudes, $\pm 30\degree$ from the equator. This would give values 
$0.5 \la h \la 0.75$
yielding an $\alpha$ in the range ${\mathbf\krange{}}$, or in terms of rotational shear, 0.05 rad d$^{-1} \la \Delta \Omega \la$ 0.07 rad d$^{-1}$ For comparison, we have listed the previously derived values of the differential rotation coefficient $\alpha$ together with our estimate in Table \ref{differential}.

\begin{table}
\caption{The strength of the differential rotation from different papers.}
\centering
\begin{tabular}{cc}
\hline
\hline
Paper & $\alpha$\\
\hline
\citet{henry1995}& $\mathbf{\pm0.038\pm0.002}$\\
\citet{kovari2007b} & $-0.0021\pm0.005$ \\
\citet{kovari2007c} & $-0.0022\pm0.0016$ \\
This paper &$\mathbf{\krange{}}$\\
\hline
\end{tabular}
\label{differential}
\end{table}

It is also of interest, how the amount of differential rotation relates to other stellar parameters in spotted stars. \citet{henry1995} give a relation for the rotation period and the differential rotation. In comparison to their result, even our quite large estimate for the differential rotation is in the expected range for this photometric period. \citet{collier2007} give a relation between the effective temperature of the star and differential rotation rate $\Delta \Omega$,
\begin{equation}
\Delta \Omega = 0.053\Bigg(\frac{T_{\rm{eff}}}{5130\rm{K}}\Bigg)^{8.6},
\label{cameron_drot}
\end{equation}
where $T_{\rm{eff}}$ is the effective temperature of the star in Kelvins and $\Delta \Omega$ is given as radians per day. Using the effective temperature $T_{\rm{eff}}=4630 \rm{K}$ from \citet{kovari2001}, we get $\Delta \Omega \approx 0.022$. This is clearly lower than our estimate for the differential rotation rate. On the other hand, we also note that some of the differential rotation estimates for similar stars, which were used to derive the above relation, have values comparable to our estimate.

In order to estimate the reliability of our result, we calculated synthetic photometry using the spot-model by \citet{budding1977}. We used a two-spot model with no differential rotation, i.e. the spots rotated with a constant period $P=P_{\mathrm{orb}}$. We sampled the synthetic light curve at the same observation times as in the original data and added normally distributed noise with zero  mean and standard deviation $\sigma_{\mathrm{N}} = 0.007$, calculated from the standard deviation of the residuals $\epsilon  = y_(t_i)-\hat{y}(t_i)$ of our CPS model. The spot model parameters were calculated to correspond to the location of the active longitudes and the spot-modelling results from \citet{kovari2001} as closely as possible. For the spot latitudes $\lambda_{\mathrm{i}}$, longitudes $\beta_{\mathrm{i}}$ and radii $r_{\mathrm{i}}$  we used values $\lambda_{1} =\spLambdaI{}$, $\lambda_{2} =\spLambdaII{}$, $r_{1} =\spRI{}$, $r_{2} =\spRII{} $, $\beta_{1}=\spBetaI{}$ and $\beta_{2}=\spBetaII{}$. 

For the inclination of the star, we used value $i=60\degree$. The values we 
used for the linear limb-darkening coefficient $u$ and spot darkening fraction 
$\kappa$ were $u=\spU{}$ and $\kappa = \spKappa{}$. The spot darkening 
fraction corresponds to a temperature difference between the photosphere 
$T_{\mathrm{phot}}=4630\mathrm{K}$ and a cool spot 
$T_{\mathrm{spot}}=4030\mathrm{K}$. The linear limb-darkening coefficient for the 
Johnson V-band was calculated using the results by \citet{claret2000} and 
bilinear interpolation. The parameters used were  $T=T_{\mathrm{phot}}$, 
$\log g = 2.5$, microturbulence $v_{\mathrm{micro}}=1.0\mathrm{km/s}$ and solar 
metallicity. All these parameters are the same as used in \citet{kovari2001}.  

Analysing this synthetic photometry, we got the value $Z_{\mathrm{synth}}=\spZ{}$. 
This result clearly demonstrates, that our differential rotation estimate is affected by the long rotation period combined with poor sampling. More complex models, such as addition of a third spot, rotating with a period of $P_{3}=19\fd47$ did not change this result, neither did addition of artificial \ff{}s.

\subsection{Active longitudes}

Active longitudes are longitudes on the surface of a star that exhibit persistent spot activity. They can appear in pairs, situated on the opposite sides of the star \citep{henry1995, jetsu1996, berdyugina1998}. The presence of active longitudes in observational data is well established in many active stars and they are thought to be manifestations of non-axisymmetric dynamo modes. \sg{} has shown very persistent active longitudes throughout its whole observational history \citep{jetsu1996}. Moreover, the active longitudes on \sg{} are synchronised with the orbital period of the tidally locked binary components, whereas on many other RS CVn stars, the active longitudes have been reported to migrate linearly in relation to the orbital reference frame \citep[e.g., ][]{berdyugina1998, lindborg2011}.

\begin{figure*}
\centering
\includegraphics[width=17 cm]{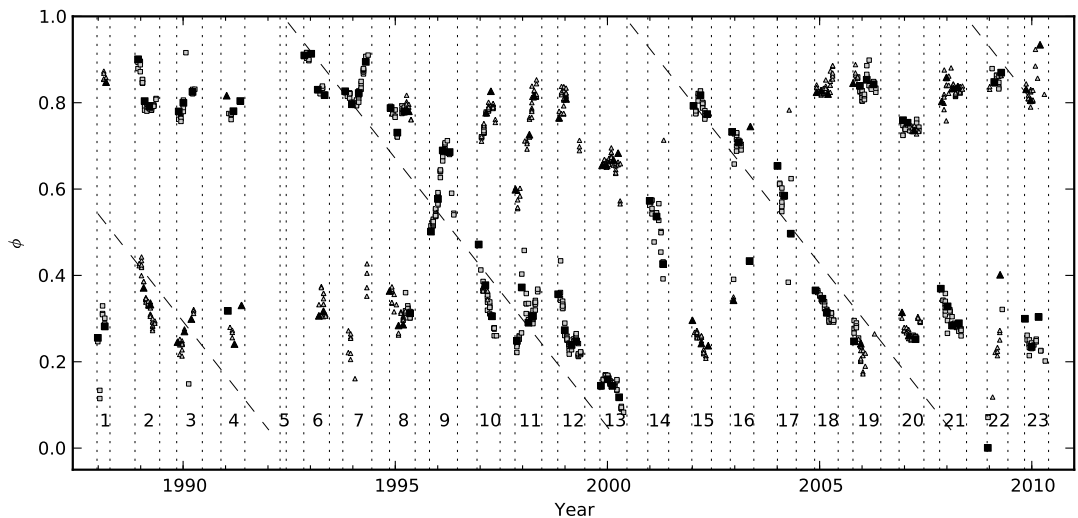}
\caption{The phases of the light curve minima of Sigma Geminorum in the orbital frame of reference. Phase $\phi=0.8$ coincides with the conjunction of the binary components, with the primary in front. The primary and secondary minima of independent datasets are denoted by black squares and triangles, respectively, grey squares and triangles are used for non-independent minima. The line is plotted using ephemeris $\rm{Yr}= 1994\fy2+8\fy0E$.}
\label{phase_diag}
\end{figure*}

The phase diagram of the light curve minima $t_{\rm{min,1}}$ and $t_{\rm{min,2}}$ (Fig. \ref{phase_diag}) shows clearly the two active longitudes, that are, with a few exceptions, present throughout the length of the whole time series. The phases $\phi_{\rm{orb}}$ were calculated with the orbital period, using the ephemeris $\rm{JD}_{\rm{conj}}= 2447237\fd02 +19\fd604471 E$ \citep{duemmler1997} and then adjusted for plotting using the formula $\phi = \phi_{\rm{orb}} - 0.2$. Thus, phase $\phi = 0.8$ marks the conjunction of the two binary components, with the primary in front. 

The epochs of the light curve minima from independent datasets, $t_{\rm{min,1}}$ and $t_{\rm{min,2}}$, were simultaneously analysed using the non-weighted Kuiper-test. The Kuiper-test is a non-parametric test that is suited for searching for periodicity in a series of time points $t_{i}$, when the phases, calculated using period $P$, $\phi_{P, i} = \frac{t_{i}}{P}\bmod 1$ have a bimodal (or even multimodal) distribution. We used the same formulation as in \citet{jetsu1996}. We used a null hypothesis ($H_{0}$), that the phases $\phi_{P, i}$ are uniformly distributed within the interval $[0,1]$, i.e., there is no periodicity present. The period search was done within the interval $0.85P_{\rm{W}}<P<1.15P_{\rm{W}}$, where $P_{\rm{W}}$ is the weighted average of the photometric periods. The resulting best period was \kpboth{} and the corresponding critical level was \kpbothQ{}.

In Fig. \ref{phase_diag} can be seen that during segments $8-13$ and $15-19$ 
the primary minima are rotating faster than the orbital period of the binary system. Therefore, we also analysed only the independent primary minima $t_{\rm{min,1}}$ using the Kuiper-test and got the best period \kppri{} with \kppriQ{}. 

If the drift of the primary minima was present only during single segments, this effect could be introduced by evolving spot pattern or differential rotation. However, the primary minima trace a clearly identifiable path, that can be seen for many years. The aforementioned effects would only apply to single spots, not whole active areas such as active longitudes.

The main contribution to $P_{\rm{min,1}}$ comes from segments SEG9, SEG14 and SEG17. During these segments the secondary minima vanish altogether and the primary minimum is shifted $\sim 0.25$ in phase. As a result, these segments with only one minima throughout the whole segment are also the only ones that are not situated near the active longitudes. This could be due to two relatively close starspots forming a single minimum, in-between their respective longitudes, or that the usual spot configuration, where spots are located near active longitudes, is temporarily disturbed by additional new starspots.

Emergence of additional star spots could also be explained with an azimuthal dynamo wave moving across the star \citep{krause1980,cole2013}.
What makes this possibility even more interesting, is that it appears that the intermittent disappearance of the stable active longitudes is somehow connected to the jump of activity between the two active longitudes; during several occasions, the primary and secondary minima switch places when the migrating primary minimum reaches either of the active longitudes. The diagonal dashed line in Fig. \ref{phase_diag} shows the movement of the primary minima. The line is plotted using ephemeris $\rm{Yr}= 1994\fy2+8\fy0 E$. 

\subsection{Flip-flops and flip-flop like events}
Flip-flops are a name coined by \citet{jetsu1993b} in their analysis of the active giant FK Com photometry. A flip-flop is an event where the primary and secondary minima suddenly switch their places in a phase diagram. We refer to these type of events as \ff{}s. An obvious interpretation is that during a \ff{}, the activity jumps from one active longitude to another. 

In the \sg{} data, several jumps in the phase diagram can be seen. Since all of these flip-flop candidates are not necessarily similar, physically related phenomena, we use the following criteria to distinguish true ff-events from apparent ones:

\begin{description}
\item{$\rm{C_{I}}$: \emph{The region of main activity shifts about 180 degrees from the
old active longitude and then stays on the new active longitude.}}
\end{description}

and 

\begin{description}
\item{$\rm{C_{II}}$: \emph{The primary and secondary minima are first separated by
about 180 degrees. Then the secondary minimum evolves into a
long–lived primary minimum, and vice versa.}}
\end{description}

Although multiple phase shifts can be found in the data, there is only one activity shift, between segments one and two, that fulfils these two criteria. In addition, there are multiple events similar to \ff{}s, that are abrupt, but not persistent, or are associated with gradual migration of the primary minima that take years to complete. These events are called \ab{}s and \gr{}s, respectively.

\subsubsection{Ff-event 1988-1989}

The only \ff{} in the data that fulfils our criteria $\rm{C_{I}}$ and $\rm{C_{II}}$ occurs between segments SEG1 and SEG2 and has been identified in previous papers that used contemporaneous data, i.e., \citet{jetsu1996} and \citet{berdyugina1998}. The first signs of a coming \ff{} can be seen at the end of SEG1, where the light curve shows a gradual deepening of the secondary minimum. The switch could have actually occurred even before the end of SEG1; in the last few models of the segment, the minima have already switched places, but the models are not reliable due to the small number of observations. At the beginning of SEG2, the previous secondary minimum has become the new primary minimum. After the \ff{}, the primary and secondary minimum stay stable for several years.

\subsubsection{Gr-events}

Signs of a coming phase shift can be seen already at the end of SEG8. In this case, however, the phase shift is gradual and caused by weakening of the primary minimum, rather than by deepening of the secondary i.e., the activity does not "jump" from an active longitude to another, rather it diminishes on one and stays constant on the other. During this segment, the star is at its brightest and correspondingly the light curve amplitude is very low, decreasing throughout segments SEG8--SEG10. 

During SEG9 the secondary minimum vanishes altogether, giving way to a relatively unstable primary, situated halfway between the previous two active longitudes. 

As seen in the light curves (Fig. \ref{lightcurves}), the new minimum is also extremely wide, most likely consisting of several large starspots. The minimum persists until the beginning of SEG10. There are several unreliable datasets in the beginning of this segment, during which two separate minima emerge from the previous single minimum. This kind of phase diagram is expected when two close starspots rotate with different periods, gradually moving away from each other \cite[][Fig. 3]{lehtinen2011}.

The Doppler images by \citet{kovari2001} taken between 1 November 1996 -- 9 January 1997, and overlapping with the first part of segment 10, show three large starspots distributed almost at equal distances in longitude on a latitude band situated between $0\degree$ and $60\degree$. The photometric spot models in the same paper also show three spots (Spots 1-3), of which Spots 1 and 2 correspond to the persistent active longitudes. This view does not support the idea of the two large spots rotating at different rates.

What we consider more likely, is that the previously almost band-like spot distribution is vanishing, or the third spot (Spot 3) slowly migrates and merges with the other active longitude (Spot 1). In our analysis, the beginning of SEG10 shows two minima, with the primary minima linearly migrating from $\phi = 0.5 $ towards the other active longitude at $\phi=0.3$. This supports the idea of the first and third spot merging. After this, the stable active longitudes again start to dominate the light curve, only this time, the primary and secondary minima have switched their places. 

During 2000-2002 and 2003-2005, the star shows again similar behaviour. During SEG13 the two active longitudes are present, but they disappear in SEG14 and are replaced by a single minimum, located halfway between the active longitudes. In SEG15, the two active longitudes are present again, and the minima have switched their places.

In SEG17 the active longitudes disappear once again and are recovered in the following segment, now with switched primary and secondary minima. Unlike during the 1996-1997 event, the amplitude of the light curve and the mean brightness of the star do not show any changes or patterns during the time span between 2000 and 2005. There is a small linearly growing trend in the mean brightness, but it does not correlate in any way with amplitude or period during that time.

\subsubsection{Multiple \ab{}s between 2005 and 2010}

The area of main activity jumps multiple times from one active longitude to another during the last five years of the time series. Unlike before, these phase shifts happen in brief succession, with intervals of only a year or two. The first \ab{}, during SEG19, can be seen in the light curves and persists until the following SEG20, where the primary minimum switches back. In SEG20 and SEG21 the same happens again: the activity jumps from one active longitude to another and back again in relatively short time. Although the phase shifts are not persistent, the primary minima are quite deep in each case and the shifts are likely to be real and not just random fluctuation or observational errors.

\subsection{Flip-flop event cycles}

We determined epochs for each ff-, ab-, and gr-event. In the case of ff- and \ab{}s, we used the mean epoch of the two primary minima between which the phase shift occurred. For \gr{}s, we determined the epoch from the moment when the primary minimum reached and stayed on the active longitude in question. \citet{jetsu1996} determined the width of the active longitudes to be 0.2 in phase, therefore we required that the primary minimum satisfied inequality $\mid\phi_{\rm{min},1}-\phi_{\rm{al},i}\mid<0.1$, where $\phi_{\rm{al},i}$ is the mean phase of an active longitude.

The mean phases of both of the active longitudes were calculated from independent datasets so that each minimum contributes only to the mean of the active longitude it is closest to in phase, and only those datasets which have two minima are used. The phases we get for the two active longitudes are $\phi_{\rm{al,1}}=0.79$ and $\phi_{\rm{al,2}}=0.30$.

The events are tabulated in Table \ref{flip-flops}. There is also one earlier flip-flop, found by \citet{jetsu1996} and later confirmed by \citet{berdyugina1998}. The epoch of the flip-flop is taken from the latter. 

\begin{figure} 
\resizebox{\hsize}{!}{\includegraphics{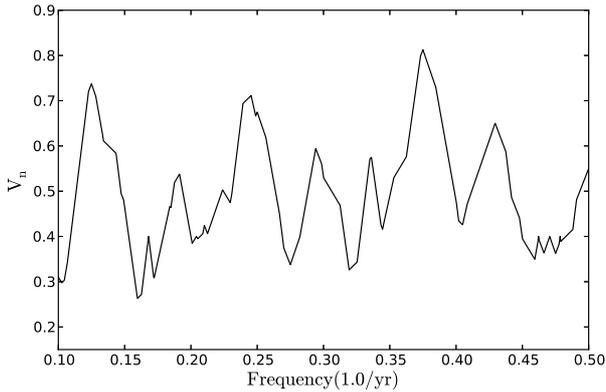}}
\caption{Kuiper-test periodogram for the ff- and gr-event epochs in Table \ref{flip-flops}.}
\label{ff_periodogram}
\end{figure}

To investigate the possibility that an azimuthal dynamo wave could be responsible for the ff- and gr-events, we analysed their respective epochs using the Kuiper-test with the period interval $P_{\rm{min}} = 2.0$ yr and $P_{\rm{max}} = 10.0$ yr. We found the best period $P_{\rm{ff,1}}=2.67\rm{yr}, \rm{V_{n}}=0.81$. We, however, consider the second best period $P_{\rm{ff,2}}=7.99$ yr, $V_{n}=0.74$ to be more plausible, partly because the shorter period would imply that there are multiple unobserved flip-flop epochs, and partly because the 2.67 year period is an integer part of the 7.99 year period. The periodogram is plotted in Fig. \ref{ff_periodogram}. The small number of time points makes it impossible to calculate any meaningful significance estimates for the Kuiper-test statistics $\rm{V_{n}}$, for this, more events would be required. 

\begin{table}
\caption{The ff-, gr- and ab-events found in the data. The first column is the number of the event, second the segment or the range of segments during which the event occurred. The next two columns give the estimated epochs of the event in JD and years. The fourth column gives the phase calculated with the ephemeris $\rm{J_{ff}}= 1981.00 + 7.99E$. The last column gives the type of the event, ff=flip-flop, gr=gradual phase shift, ab=abrupt phase shift.}

\centering
\begin{tabular}{lccccc}
\hline
\hline
Event &Segment & HJD& year& $\phi_{\rm{ff}}$& Event type\\
\hline
1 & --     &  --       & 1981.0 & 0.00& ff \\
2 & 1--2   & 2446331.0 & 1988.5 & 0.94& ff \\
3 & 8--10  & 2450451.1 & 1997.0 & 0.00& gr  \\
4 & 13--15 & 2452255.6 & 2001.9 & 0.62& gr  \\
5 & 16--18 & 2453305.9 & 2004.8 & 0.98& gr  \\
6 & 19     & 2453676.2 & 2005.8 & 0.11& ab  \\
7 & 20     & 2454250.7 & 2007.4 & 0.31& ab  \\
8 & 21--22 & 2454673.6 & 2008.6 & 0.45& ab  \\
9 & 22--23 & 2454908.2 & 2009.2 & 0.53& ab  \\
\hline
\end{tabular}

\label{flip-flops}
\end{table}

\section{Discussion}

\begin{figure*}
\centering
\includegraphics[width=17 cm]{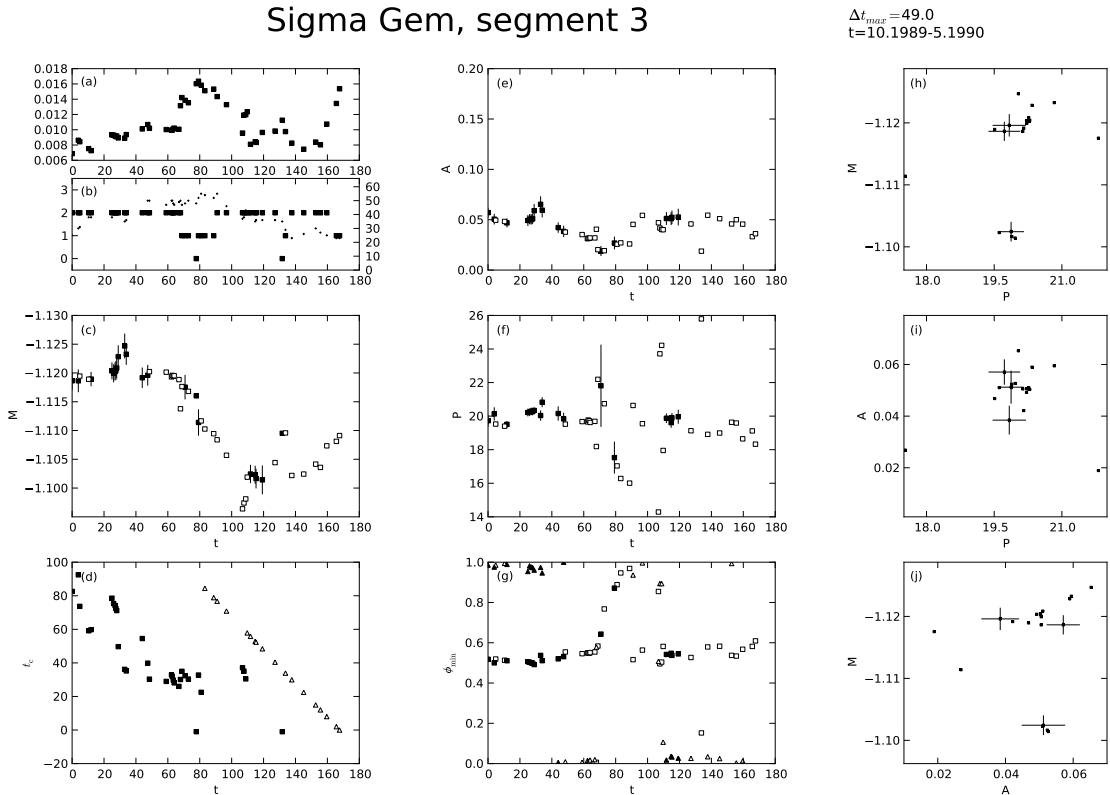}
\caption{CPS-analysis of segment SEG3. The contents of the panels are explained at the end of Sect. 3.}
\label{seg3}
\end{figure*}
\subsection{Differential rotation}

The value we get for differential rotation is an order of a magnitude greater than the values found in previous studies. The main culprit is most likely the long rotation period. In some cases, this leads to poor phase coverage which in turn leads to large uncertainties in the period estimates. Another problem the long rotation period causes is the possibility that the spot structure changes on the surface of the star. In the phase diagrams of some datasets this can be seen as a superposition of two light curves with noticeably different shapes. In a worst case scenario these two effects appear simultaneously, i.e., the spot structure changes between successive rotations, but this is not noticeable due to poor phase coverage. Thus the phase diagram can create an illusion of a unique, continuous light curve, while in fact, it was created by two different spot configurations, introducing an error to the period estimate. 

\citet{henry1995} used some of the same data analysed in this paper. The discrepancy between our and their differential rotation estimates can be easily attributed to the different approaches that were used. \citet{henry1995} used spot modelling and the differential rotation estimate was derived from the maximum and minimum periods determined from the spot migration curve of each spot. 

In our analysis even slight changes in the spot structure, occurring faster than the rotational 
period, could lead to a change in the period estimate, which we then interpret as differential rotation. 
In \citet{lehtinen2011}, only the signal-to-amplitude ratio was considered when the amount of spurious period change was estimated. 
Our simulated data indicated that also the sampling effects are a considerable source of spurious period changes and can result in overestimates of the differential rotation.

When comparisons are made to other stars with comparable period, the range of differential rotation these stars exhibit is greater than the difference between measurement techniques. There is also considerable doubt if starspots are even reliable proxies of differential rotation. Even spots on the same latitude might have different migration rates due to different anchor depths. This is shown by recent numerical simulations indicating that if
observed starspots are caused by a large-scale dynamo field, their movement is not necessarily tracing the surface differential rotation, but the movement of the magnetic field itself \citep{korhonen2011}. Finally, observed spots are not necessary even stable or might consist of many small rapidly evolving starspots instead of one large spot. In any case, it is clear that the use of the CPS-method might greatly overestimate differential rotation, if the rotation period is long.

\subsection{Active longitudes and flip-flop events}

We find three kinds of events in which the activity moves from an active longitude to another. In addition to the single flip-flop fulfilling the criteria $\rm{C_{I}}$ and $\rm{C_{II}}$ (\ff{}s), we find abrupt (\ab{}s) and gradual phase shifts (\gr{}s). It is not clear whether or not these different kind of events are caused by the same phenomenon or not. A similar kind of two-fold behaviour of ab- and gr-events has been reported in FK Com, by e.g. \citet{olah2006} and \citet{hackman2013}.

The ab-events are similar to the flip-flops discovered by \citet{jetsu1993b}: sudden shifts of primary minima from one active longitude to another. The only difference is that in \ab{}s, the phase-shift is not persistent and the area of the main active region shifts back to the original active longitude, roughly after a year or so. Unfortunately, it is almost impossible to determine which \ab{}s are fundamentally different from "real" \ff{}s, i.e., sudden, but persistent phase shifts. Some of the \ab{}s could fail to fulfil the criteria for \ff{}s, simply because the event is followed by an unrelated \ab{}, creating an appearance of a non-persistent phase-shift.

During the \gr{}s, the location of the main active region shifts by $\sim 90 \degree$ in longitude and the photometry shows only one wide minimum. The disappearance of the two long-lived active longitudes and their replacement with only one minimum could be illusory at least in SEG9. The overlapping Doppler images dated to the beginning of SEG10 show that there is a large spot area near the longitude the single minimum was located at.

The appearance of spots between the active longitudes resembles what \citet{olah2006} found in photometry of FK Com, and called \emph{phase-jumps}. In a \emph{phase-jump}, old active areas disappear and then new ones emerge, with an offset of roughly $90\degree$ in respect to the original active longitudes. In FK Com, the phase jumps cause the active longitudes to stay displaced for a much longer time \citep{hackman2013}. It could be that the binary nature of \sg{} affects the preferred location of the active longitudes and this displacement is not long lasting.

In segments SEG10, SEG14, SEG17 and SEG18 the primary minimum traces a path towards the active longitude situated at $\phi_{\rm{al,2}}=0.3$. This may imply that in addition to the stable non-axisymmetric dynamo mode, there is also a possible azimuthal dynamo wave present, rotating faster than the star itself. This is also suggested by the period of the primary minima, which is shorter than the orbital period of the tidally locked binary system.

If present and rotating at a constant rate, the wave would return to a same 
active longitude every 7.9 years. This period is remarkably close to the 7.99 
year period we find from the epochs of the gr- and ff-events. It is possible 
that at least some of those observed events occur when a spot structure 
corresponding to a moving dynamo wave interferes with the stable active 
longitudes, either strengthening or weakening the minima as it passes by. This 
does not prevent the \ab{}s from also being caused by this mechanism. In this 
model, the time between successive flip-flops is equal only in the case that 
the apparent spot coverage on both of the active longitudes is equal. If the 
spot coverage on either of the active longitudes is greater, the primary 
minimum will stay on this active longitude for a longer time. To further 
complicate things, the spot coverage on the active longitudes may also change 
independently from the migrating active area, and this can prevent the 
observation of the gr-events.

If there is an active region migrating in the frame of the orbital period, we should see periodic variation in $A$. The interference between the stationary
active longitudes and the migrating region should modulate the light curve with an amplitude envelope with the same period as the flip-flop event cycle. We detect no clear sign of such modulation, although events 3 and 4 in Table \ref{flip-flops} are associated with relatively low values of $A$. 

It is possible that this amplitude effect is masked by short term spot evolution. This seems plausible, since the variation in $A$ between independent datasets within one segment is quite large. Intriguingly, there are also disturbances in the light curves at multiple occasions, at the same epochs when the presumed dynamo wave passes an active longitude. An example of this can be seen in SEG3 (Fig. \ref{seg3}), where these abrupt changes in the light curve even prevent reliable CPS modelling. Similar behaviour can be seen in segments SEG11 and SEG19. As for the other  CPS-parameters, $M$ and $P$, there seems to be no obvious connection between them and the ff-, gr- and ab-events. Also the periods found from these parameters are different from the Kuiper-test periods. On the other hand, one may speculate that the 2.67 and 8.5 year periods found by \citet{strassmeier1988} and \citet{henry1995}, respectively, are somehow connected to the 2.7 and 8.5 year Kuiper-test periods. If either of those periods is real and due to a migrating spot area, this could very well be reflected in the mean brightness of the star.

\section{Summary and conclusions}

By applying the CPS method to photometry of \object{$\sigma$ Gem} we have been
able to study in detail the long-term evolution of the mean brightness ($M$), 
light curve amplitude ($A$) and, photometric minima ($t_{\rm{min,1}}, 
t_{\rm{min,2}}$) and photometric rotation period ($P$) of the star. The best
periodicities in $M$, $A$ and $P$ were: \Pmag{} yr, \Pamp{} yr and \Pper{} yr. 
Variations in $P$ could be explained by differential rotation, for which
we estimated a coefficient of $\krange{}$. From the combined time 
point series of the both primary and secondary minima, we found a period 
\kpboth{}. When only primary minima were analysed, we retrieved the period 
\kppri{}. Furthermore we analysed flip-flops as well as other gradual and 
abrupt phase shift events and found that the best period for these would be
7.99 years. However, the small amount of events prevented a meaningful 
significance estimate this period.

There appears to be no direct connection between the periods found from the $A$, $M$ and $P$. The only obvious connection between $P$ and the other model parameters could be due to differential rotation. The differential rotation estimate we get from the period 
changes
is extremely large when compared to previous  analyses \citep{kovari2007b,kovari2007c}.
Using synthetic photometry, we demonstrate that this is at least partly due to the long rotation period of the star, which sometimes leads to sparse and uneven phase coverage. This can cause large fluctuations in the period estimates, and thus, an unreasonably large differential rotation estimate.

We also confirm the presence of previously found persistent active longitudes, which are tied to the orbital reference frame of the binary system. The most interesting result we present in this paper is the possible connection between the flip-flop like events and the drift of the primary minima. This may imply that there is a superposition of two dynamo modes operating in the star. One would be tied to the orbital period of the binary system, while the other one could manifest itself as an azimuthal dynamo wave, rotating faster than the star.
Signs of such dynamo waves have been observed in other stars 
\citep[][e.g.]{lindborg2011,hackman2011,hackman2013} and have also been 
reproduced in numerical MHD-simulations \citep{cole2013}.

Such an
azimuthal dynamo 
could disturb the stable active longitudes present in the star, creating the ff-, gr- and ab-events. In order to find supporting evidence for the presence of 
a 
propagating
dynamo wave, it would be necessary to obtain new Doppler images of the star, preferably at least two sets taken at different times, and check, whether or not there are star spots in areas indicated by the ephemeris $\rm{Yr} = 1994\fy2+8\fy0E$.

\begin{acknowledgements}
This work has made use of the SIMBAD data base at CDS, Strasbourg, France and 
NASA’s Astrophysics Data System (ADS) bibliographic services. The work by PK 
and JL was supported by Vilho, Yrjö and Kalle Väisälä Foundation. The work of 
TH was financed by the project "Active stars" at the University of Helsinki. 
The automated astronomy program at Tennessee State University has been 
supported by NASA, NSF, TSU, and the State of Tennessee through the
Centers of Excellence program.
We thank the referee for valuable comments on the manuscripts.
\end{acknowledgements}
\bibliographystyle{aa}
\bibliography{references}
\end{document}